# Learning to Dock: Geometric Deep Learning for Predicting Supramolecular Host-Guest Complexes

Zidi Wang[1,2#], Tao Zhang[3#], Muyao Yu[1,2], Chuyi Zhou[1,2], Zezhao Xu[1,2], Huiyu Liu[1,2], Yuzhen Wen[1,2], Linjiang Chen[4*], Jie Zheng[3*] and Shan Jiang[1,2*]

[1]School of Physical Science and Technology, Shanghaitech University, Shanghai 201210, China

[2]State Key Laboratory of Advanced Medical Materials and Devices, ShanghaiTech University, Shanghai 201210, China

[3]School of Information Science and Technology, ShanghaiTech University, Shanghai 201210, China

[4] State Key Laboratory of Precision and Intelligent Chemistry, University of Science and Technology of China, Hefei, Anhui 230026, China.

[#] These authors contributed equally.

e-mail: linjiangchen@ustc.edu.cn; zhengjie@shanghaitech.edu.cn; jiangshan@shanghaitech.edu.cn

## Abstract

Predicting non-covalent host–guest recognition remains challenging due to the complex interplay of electrostatics, dispersion, and steric effects, and the limited transferability of existing docking approaches to synthetic supramolecular systems. Here we present *DeepHostGuest*, a geometric deep-learning framework that learns generalizable recognition principles directly from experimentally resolved host–guest structures. Hosts are encoded as electrostatic surfaces and guests as molecular graphs, enabling transferable learning across diverse supramolecular systems. *DeepHostGuest* achieves high-accuracy predictions (RMSD ≤ 2 Å for 80.8% of test cases), substantially outperforming classical docking without case-specific tuning. Notably, the method generalizes beyond its training domain to crystalline sponge systems, accurately capturing the binding of large amphiphilic molecules within metal–organic cages. Beyond predicting binding conformations, the structures generated by *DeepHostGuest* serve as a reliable basis for accurate binding free-energy calculations. Density Functional Theory (DFT)-calculated affinities correlate well with experiment, enabling structure–property relationships across 876 host–guest complexes spanning 34 host families. Interpretable feature analysis reveals that binding strength arises from a cooperative interplay of host polarity, guest hydrophobicity, and geometric complementarity, with distinct design regimes across supramolecular classes. Together, these results establish data-driven molecular recognition as a practical route to predictive supramolecular design, enabling high-throughput virtual screening and rational optimization of functional host–guest systems.

**Introduction**

Biological systems achieve remarkable selectivity in molecular recognition through sophisticated spatial and electrostatic engineering.[1] By confining substrates within precisely defined binding pockets, nature controls chemical reactivity, regulates transport, and enables exquisite molecular discrimination.[2] This elegant principle—containment as control—has inspired decades of synthetic supramolecular chemistry aimed at engineering artificial hosts that replicate biological functions.[3] For example, cyclodextrins can enhance drug solubility and bioavailability by orders of magnitude;[4] cucurbiturils display protein-like binding selectivity;[5] and coordination cages accelerate reactions by pre-organizing reactive intermediates within well-defined cavities.[6] These successes demonstrate the transformative potential of engineered supramolecular environments for controlling molecular behavior. Yet this promise remains substantially unrealized: the discovery of genuinely new functional hosts and identification of optimal guests for specific applications continues to rely overwhelmingly on empirical trial-and-error synthesis and screening.[7, 8] This empirical method, while productive, is slow, resource-intensive, and fundamentally limits the breadth of chemical space that can be practically explored.

The core challenge is predicting the binding conformation—the spatial arrangement that governs non-covalent interactions, determining affinity, selectivity, and reactivity.[7, 9] This problem shares surface similarities with protein–ligand docking, which has transformed drug discovery through decades of algorithmic refinement and parameterization.[10] However, the supramolecular context presents distinct and formidable challenges. First, synthetic hosts span chemical and structural diversity: the same host family—cucurbiturils—ranges from CB[5] with a rigid, hydrophobic cavity to CB[10] with expanded dimensions;[5] cyclodextrins exist as distinct isomers (such as α, β, γ) with different cavity sizes and rigidities;[4] organic cages exhibit diverse geometries and accessible cavities;[11-13] metal–organic cages involve three-dimensional (3D) architectures with varying topology and metal–ligand chemistry.[6, 14] In addition, binding mechanisms span a spectrum, with hydrophobic encapsulation dominating in some systems, electrostatic interactions in others, and balanced contributions from multiple interaction types in many.[15, 16] Existing computational strategies capture parts of this complexity but still face challenges in general applicability. For example, *stk* provides a flexible framework for constructing host–guest structures. [17] However, it relies on user-defined guest placement and optimization, limiting automation. In contrast, *cgbind* offers a more automated workflow for metal–organic cages by positioning guests near the host centroid and estimating interactions using empirical scoring functions,[18] but its scope is largely restricted to metal–organic systems. More generally, protein–ligand docking algorithms can, in principle, be adapted to supramolecular host–guest recognition [19-21], yet their transferability is constrained by biomolecule-oriented input representations, scoring functions, and parameterizations.[19, 22, 23] As a result, many chemically relevant features of host–guest systems are inadequately captured. In particular, supramolecular complexes frequently incorporate elements such as lanthanides and transition metals that fall outside standard force-field parameterizations, hindering the accuracy and reliability of conventional docking approaches for these systems. What is needed is a general framework capable of learning binding principles across diverse host chemistries from experimental structures, without requiring system-specific tuning.

In this study, we introduce *DeepHostGuest*, a geometric deep learning (GDL) framework that couples electrostatic potential representations of hosts with molecular graph descriptions of guest molecules, enabling predictive supramolecular recognition. Recent advances in GDL offer a compelling alternative to traditional docking and scoring approaches. Unlike classical methods based on hand-crafted features, GDL models learn directly from 3D molecular geometries and connectivity, extracting hierarchical representations from raw structural data. Proof-of-concept studies show that GDL outperforms conventional quantitative structure–activity relationship models in molecular property prediction and achieves competitive accuracy in protein–ligand docking. These advances underscore a central insight:

3D molecular geometry encodes rich chemical information that can be extracted more flexibly and effectively by data-driven models than by manually designed scoring functions. Supramolecular host–guest systems represent an especially attractive domain for this approach. Compared with proteins, host–guest pairs explore far broader and more chemically diverse design spaces, while many hosts feature well-defined, geometrically simple cavities that enable direct analysis of recognition. Moreover, the availability of large, curated experimental datasets—most notably crystallographically resolved host–guest complexes in the Cambridge Structural Database (CSD)[24]—creates a unique opportunity to develop a general-purpose learning framework for supramolecular recognition.

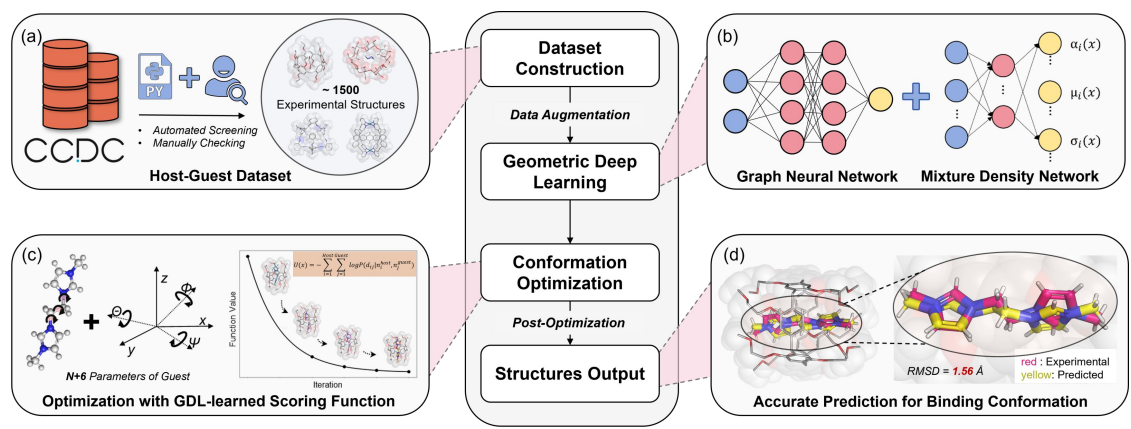

**Scheme 1.** Pipeline of the *DeepHostGuest* framework for host–guest binding prediction. (a) Dataset: 1,499 experimentally resolved complexes augmented tenfold via random rotations and translations. (b) Geometric deep learning (GDL): Graph neural networks combined with mixture density networks learn a universal scoring function capturing transferable recognition motifs. (c) Optimization: Scoring function guides conformational and positional sampling with post-refinement. (d) Structural output. *DeepHostGuest* outputs binding conformations that closely reproduce experimental structures, demonstrating its predictive accuracy.

Trained on 1,499 experimentally resolved complexes with further 3D data augmentation, *DeepHostGuest* learns a transferable scoring function that enables automated prediction of host-guest binding conformations across diverse supramolecular systems (Scheme 1). The framework reproduces experimental binding geometries with high fidelity, achieving root-mean-square deviations (RMSD) of $\leq$ 2.0 Å for 80.8% of test cases, and consistently outperforms classical docking approaches without any system-specific parameterization. Notably, *DeepHostGuest* exhibits strong out-of-distribution generalization, accurately predicting binding modes in crystalline sponge systems containing large, flexible, and amphiphilic guests far beyond the training domain. Beyond predicting host–guest conformations, the *DeepHostGuest*-generated geometries provide reliable inputs for binding free-energy calculations, enabling systematic structure–activity relationship analysis across 876 experimentally characterized complexes. SHAP (SHapley Additive exPlanations)-based feature attribution further reveals host-class-specific recognition principles, translating statistical learning outcomes into actionable molecular design rules. Collectively, these results demonstrate that supramolecular recognition can be learned directly from experimental structures, establishing a scalable framework that advances the field from empirical trial-and-error towards predictive, data-driven supramolecular discovery.

## Results and Discussion

### Dataset Construction and Model Architecture

The performance of GDL models depends critically on the availability of large, diverse, and high-quality structural datasets. Whereas protein–ligand modelling benefits from well-established repositories such as the Protein Data Bank,[25] no analogous resource exists for supramolecular host–guest assemblies. To bridge this gap, we developed a dedicated data-mining and curation pipeline that systematically extracts experimentally resolved host–guest complexes from the CSD. The pipeline comprises four stages (Figure 1a, see Supporting Figure S1 and Table S1 in Section 1 for details). Crystallographic data are automatically parsed to identify molecular components, which are then extracted and classified as hosts or guests. Geometric and intermolecular criteria are applied to identify valid inclusion complexes, followed by manual curation to remove artefacts and misassignments. This procedure yielded a curated set of 1,499 high-confidence host–guest complexes spanning macrocycles, organic cages, and metal–organic cages, with system sizes ranging from tens to several hundred atoms (see Supporting Figure S2-S4). The resulting chemical and structural diversity enables the model to learn transferable recognition patterns across supramolecular space. To enhance robustness, each complex was subjected to 3D spatial augmentation through random rotations and translations, generating multiple augmented structures per entry (see Supporting Figure S5). This augmentation enforces 3D spatial transformations, such as rotations and translations, effectively expanding the dataset and improving model generalization.[26] The final dataset supports accurate conformational prediction for previously unseen host–guest systems and provides a reliable structural basis for modelling molecular recognition.

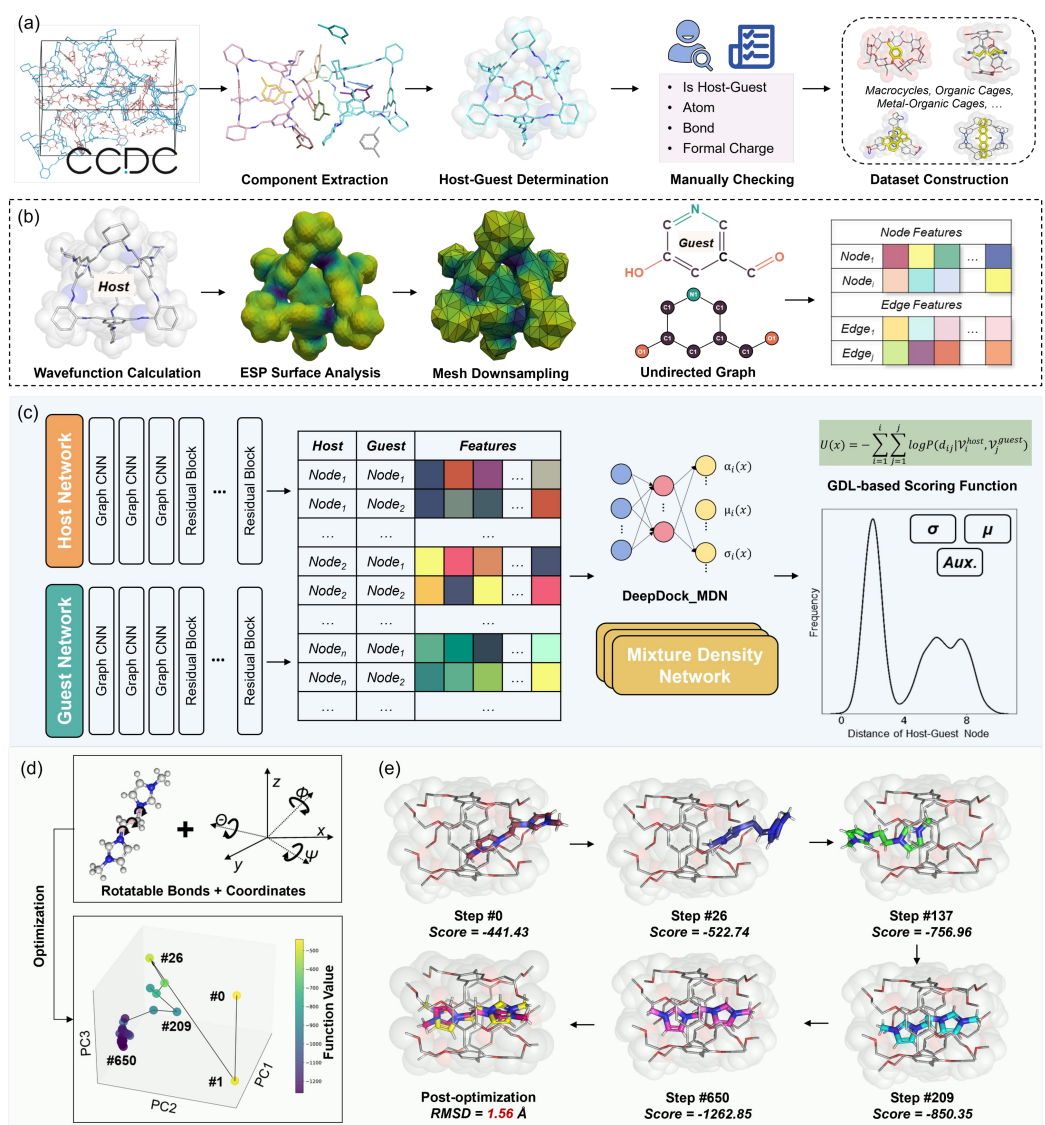

**Figure 1.** Dataset construction, model architecture, and binding optimization workflow of *DeepHostGuest*. (a) Automated curation pipeline for extracting host–guest complexes from crystallographic databases, including molecular identification, host–guest classification, geometric filtering, and manual validation, yielding a diverse set of supramolecular systems. (b) Dual molecular representation scheme. Hosts are encoded as polygonal surface meshes annotated with electrostatic potential (ESP), preserving cavity geometry and electrostatic landscapes, while guests are represented as molecular graphs capturing atomic and bond-level chemical features. (c) GDL architecture of *DeepHostGuest*. Host and guest representations are processed by separate graph convolutional

networks (*HostNetwork* and *GuestNetwork*), whose features are pairwise combined and passed to a Mixture Density Network (MDN) adapted from *DeepDock* to learn probabilistic distance distributions between host surface points and guest atoms, yielding a GDL-based scoring function. (d) Conformational search and optimization scheme, in which guest rotational degrees of freedom and internal torsions are optimized in the learned scoring landscape. (e) Representative optimization trajectory illustrating progressive refinement of the guest binding pose within the host cavity, converging to an optimal conformation with sub-angstrom accuracy relative to experiment.

From the curated structural dataset, host molecules are represented as polygonal surface meshes annotated with electrostatic potential (ESP) values (Figure 1b; Supporting Figure S6 in Section 2), capturing both 3D geometry and electrostatic features that govern molecular recognition. To ensure computational efficiency, high-resolution meshes are systematically downsampled while preserving essential geometric and electrostatic patterns required for accurate binding prediction. Guest molecules are encoded as molecular graphs, with atoms as nodes and bonds as edges. Node features describe atomic properties such as element type and formal charge, while edge features capture bond type, aromaticity, and conjugation. This graph-based encoding is widely adopted for representing small molecules, as it compactly captures both connectivity and chemical attributes, making it well-suited for deep learning tasks.[27] This dual representation strategy reflects the complementary roles of hosts and guests in supramolecular recognition. Host ESP surfaces define molecular structures, cavity geometries, and electrostatic distributions that constrain binding, whereas guest graphs provide an efficient representation by preserving their atomic and bond-level chemical features. Together, they provide a comprehensive representation of the physicochemical determinants of host–guest complexes.

The processed host and guest representations are fed into the GDL model, which integrates a Graph Neural Network (GNN) with a Mixture Density Network (MDN) adapted from the *DeepDock* framework (Figure 1c);[27] the GNN serves to compress graph-based molecular information, while the MDN learns the probabilistic distance relationships between host and guest nodes.

The GNN consists of two subnetworks: the *HostNetwork* processes ESP meshes, and the *GuestNetwork* encodes molecular graphs. Each employs a three-layer graph convolutional backbone followed by 10 residual layers, enabling efficient message passing across nodes and extraction of both local and global structural features. Outputs from the two networks are combined through pairwise concatenation, establishing correlations between guest molecular graph nodes and host ESP surface points. These integrated features are then passed to the MDN module, which predicts probability distributions $P(d_{ij}|v_i^{host}, v_j^{guest})$ for each host–guest node pair, where $d_{ij}$ denotes the distance between the $i$-th guest node and the $j$-th host node. A global scoring function $U(x)$ is then constructed as the negative log-likelihood summed over all pairs, such that minimizing $U(x)$ corresponds to maximizing the probability of the predicted binding conformation under the learned model (see Supporting Figure S7 for training details).

Once trained, the GDL-derived scoring function $U(x)$ is used to predict unknown host–guest binding conformations. Given a host–guest pair, the guest's translational and rotational degrees of freedom, together with the torsional angles of its rotatable bonds, are treated as optimizable variables (Figure 1d). Binding poses are identified by minimizing $U(x)$ using differential evolution (DE), a population-based global search algorithm suited to high-dimensional conformational search.[28] Starting from random initial placements outside the host cavity, the guest iteratively migrates into the cavity through coordinated optimization of position, orientation, and internal conformation, converging to the most probable binding mode under the learned distance distributions (Figure 1e).

Because the learned scoring function encodes statistical distance preferences but does not explicitly enforce chemical constraints, occasional steric clashes may arise.[29] To address this, a penalty function $P(x)$ derived from a pseudo–Lennard-Jones potential is introduced to prevent excessively close interatomic contacts (Supporting Figure S8).[30] The refined scoring function, $U(x) + P(x)$, retains the learned probabilistic potential while enforcing steric plausibility by penalizing atomic overlaps, producing conformations that are both statistically plausible and chemically sound (Supporting Figure S9).

To further improve physical fidelity, the predicted structures are subjected to post-optimization using molecular simulations. We employ the semiempirical DFT method *GFN2-xTB*, which offers broad element coverage, balanced accuracy, and high computational efficiency and is well established for host–guest systems.[31] This final relaxation step refines local geometries, ensuring that the predicted binding conformations are physically consistent, chemically meaningful, and suitable for downstream property calculations and mechanistic analysis. The optimizations are detailed in Supporting Information Section 3.

**Binding Conformation Prediction Performance**

To assess *DeepHostGuest*'s predictive accuracy, we evaluated the model on a held-out test set of 99 host–guest complexes spanning diverse supramolecular architectures (crown ethers, pillararenes, cucurbiturils, organic cages, metal–organic cages). We compared performance against two widely adopted benchmarks: *Glide* (Schrödinger), the commercial standard in structure-based drug discovery,[19, 32-34] and *AutoDock Vina*, an open-source docking program tested with both default *Vina* and *Vinardo* scoring functions.[20, 21, 23] Prediction accuracy was quantified by root-mean-square deviation (RMSD) from experimentally determined binding conformations—a standard metric in docking literature defined as the geometric mean distance between predicted and experimental atomic coordinates after optimal rotation and translation. We adopted RMSD ≤ 2 Å as the success criterion, a threshold representing chemical accuracy sufficient for identifying dominant binding modes, and it is standard in docking literature.[35] The simulation parameters of Glide and *AutoDock Vina* are detailed in Supporting Information Section 3.

**Table 1.** Performance of *DeepHostGuest* and classical docking methods in predicting host–guest binding conformations

| Method | RMSD ≤ 2Å | 2Å < RMSD ≤ 3Å | RMSD > 3Å | Failed |
| --- | --- | --- | --- | --- |
| *DeepHostGuest* | 80.8% | 8.1% | 11.1% | 0.0% |
| *Glide* | 64.6% | 17.2% | 8.1% | 10.1% |
| *AutoDock (vina)* | 51.5% | 17.2% | 27.3% | 4.0% |
| *AutoDock (vindardo)* | 55.6% | 10.1% | 30.3% | 4.0% |

As summarized in Table 1, *DeepHostGuest* achieved high-precision predictions (RMSD ≤ 2 Å) for 80.8% of cases, compared with 64.6% for *Glide*. This demonstrates the capability of the GDL-based framework to deliver accurate predictions across diverse supramolecular systems. Both methods outperformed *AutoDock Vina*, which reached 51.5% with *Vina* and 55.6% with *Vinardo*, highlighting the challenges faced by general-purpose docking algorithms when directly applied to supramolecular systems that differ substantially from biomolecular environments.

Beyond accuracy, *DeepHostGuest* exhibited superior robustness across chemical space. *Glide* failed to generate valid poses in 10.1% of cases, while *AutoDock Vina* showed limitations for complexes containing elements such as boron or transition metals, which are common in host–guest chemistry. In contrast, *DeepHostGuest* successfully handled all systems, reflecting its architecture-agnostic design that learns directly from structural data rather than relying on predefined atom types or empirical parametrization. Together, the combination of good predictive accuracy and broad applicability establishes *DeepHostGuest* as a practical and reliable tool for high-throughput modelling and systematic exploration of supramolecular recognition.

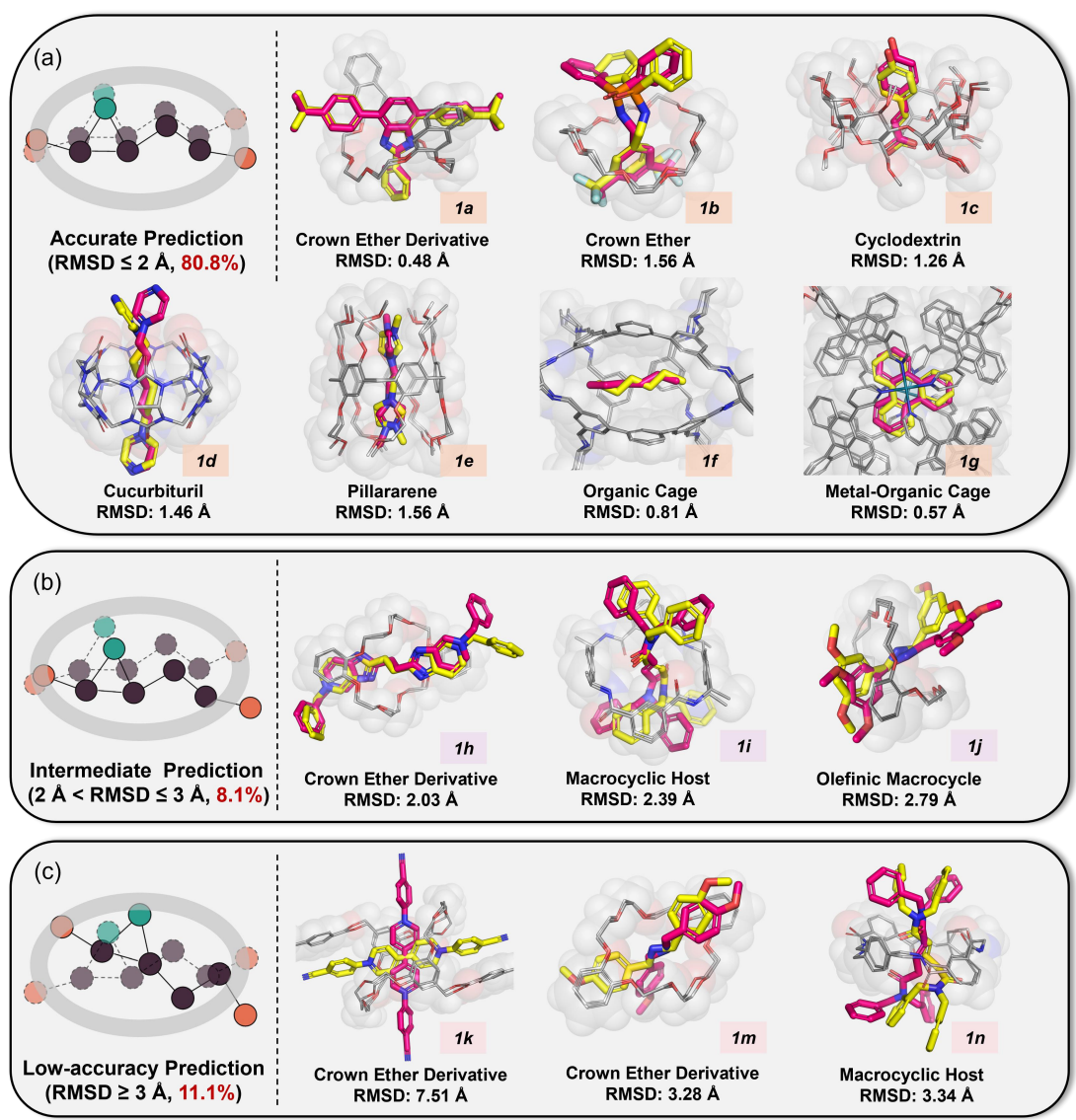

**Figure 2.** Representative DeepHostGuest predictions across RMSD regimes. (a) High-accuracy predictions (RMSD ≤ 2 Å), showing near-atomic agreement with experiment across diverse host−guest systems, including crown ethers, pillararenes, cucurbiturils, organic cages, and metal-organic cages. (b) Intermediate predictions (2 Å < RMSD ≤ 3 Å), where the correct binding region is preserved, and deviations are limited to flexible peripheral substituents. (c) Low-accuracy predictions (RMSD > 3 Å), mainly observed for macrocyclic hosts, arise from inverted binding orientations (structure 1k) or misalignment of bulky guest substituents beyond the primary binding region. These cases reflect alternative binding modes and the absence of explicit crystal-environment effects. Experimental structures are shown in red and predicted conformations in yellow; hydrogen atoms are omitted for clarity.

To further evaluate the performance of *DeepHostGuest*, we examined representative predictions across three RMSD regimes, highlighting both strengths and limitations of the model. For high-accuracy predictions (RMSD ≤ 2 Å; Figure 2a), the model achieves near-atomic agreement with experimental binding geometries across a wide range of supramolecular systems, including crown ethers, pillararenes, cucurbiturils, organic cages, and meta-organic cages (structures *1a–1g*). In these cases, *DeepHostGuest* correctly identifies binding sites and reproduces dominant recognition motifs such as hydrogen bonding, π–π stacking, CH–π interactions, and ion–dipole contacts. Achieving this level of accuracy for over 80% of systems demonstrates that the model generalizes effectively across chemically and architecturally diverse host–guest classes.

In the intermediate regime (2 Å < RMSD ≤ 3 Å; Figure 2b), predictions largely preserve the correct binding region, with deviations confined mainly to flexible substituents extending beyond the host cavity. For larger and more flexible guests, such as the rotatable phenyl group in structure *1j*, peripheral conformations vary while the core binding geometry and key interaction motifs remain intact (RMSD = 2.79 Å). These results indicate that *DeepHostGuest* provides reliable structural insight even for systems with substantial conformational freedom.

Larger deviations (RMSD > 3 Å; Fig. 2c) occur predominantly in macrocyclic hosts and account for approximately 11% of the dataset. One class of error arises from inverted binding orientations, where the guest adopts the opposite insertion direction relative to experiment while maintaining chemically reasonable local contacts, as illustrated in structure *1k* (RMSD = 7.51 Å). A second source of deviation involves bulky guests with peripheral groups extending beyond the primary binding region. These discrepancies reflect the molecular-level scope of the current model. Although trained on experimentally resolved crystal structures, *DeepHostGuest* does not explicitly account for crystal packing, lattice constraints, or solvation effects, which can stabilize specific global orientations in the solid state. As a result, the model may generate poses that are locally accurate yet globally misaligned with respect to the crystallographic reference. Incorporating approximate environmental effects and expanding conformational sampling are likely routes to further improvement.

Overall, *DeepHostGuest* accurately reproduces binding geometries and recognition modes for nearly 90% of the evaluated host–guest systems, demonstrating robust predictive capability for supramolecular recognition. The remaining challenging cases highlight opportunities for future refinement, but do not detract from the framework's utility as a practical tool for structural modelling, high-throughput screening, and the accelerated discovery of host–guest assemblies.

**Generalization to Crystalline Sponge Systems**

A stringent test of model generalization is performance on systems substantially different from the training set. We applied *DeepHostGuest* to eight crystalline sponge systems—a specialized class of porous metal–organic cages containing large, flexible, amphiphilic guest molecules absent from our training data.[16] Specifically, we examined guests encapsulated within octahedral $Pd_6L_4$ metal–organic cages featuring large hydrophobic cavities. The guest molecules range from 30 to 109 atoms and contain up to ten rotatable bonds, making them substantially larger and more flexible than typical training examples, as shown in Figure 3a.

Given only isolated host crystal structures and unbound guest molecules—without any prior knowledge of experimental host–guest complexes—we predicted binding conformations using the trained *DeepHostGuest* model. Due to the high flexibility of guest molecules, conformational search was performed before prediction and the ten lowest-energy guest conformations were fed into *DeepHostGuest* and compared with experimental structures (see Supporting Figure S10 for detailed prediction workflow in Section 4).

Results were successful: six systems were predicted with high accuracy (RMSD ≤ 2 Å), and one additional case achieved an RMSD of 2.73 Å, reproducing the experimental binding mode with only minor deviations. For rigid amphiphilic guests such as *3a* and *6* in Figure 3, the model achieved good accuracy, correctly embedding the guest within the cage cavity and capturing dominant hydrophobic interactions. For more flexible guests, including *4b* and *4c* with up to eight rotatable bonds, *DeepHostGuest* accurately reproduced the hydrophobic–hydrophilic organization, positioning the hydrophobic backbone within the cavity and orienting polar side chains toward the cage windows (RMSD = 1.75 Å and 1.89 Å, respectively). A similar behavior was observed for *3b*, which contains nine rotatable bonds and was predicted with RMSD = 2.73 Å, preserving the experimentally observed binding mode. The largest deviation occurred for guest *5a* (109 atoms), which adopts an extended conformation spanning two cage windows. While *DeepHostGuest* captured this inter-window binding pattern, it did not fully reproduce the detailed geometry, likely reflecting the limited representation of such large guests in the current training set. The detailed zoom-in structures are presented in Supporting Figure S11 in Section 4.

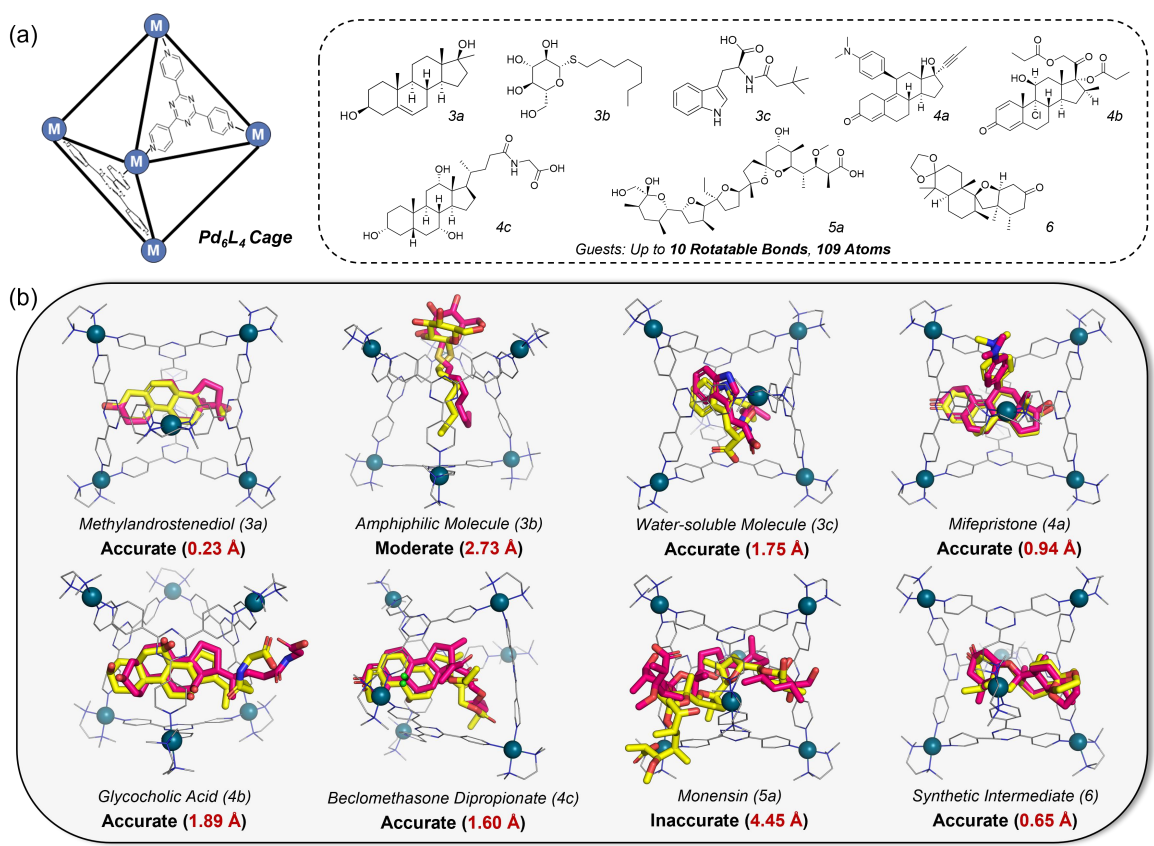

**Figure 3.** Binding conformation prediction for crystalline sponge systems. (a) Structures of an octahedral $Pd_6L_4$ metal–organic cage with a large hydrophobic cavity used as the host, together with eight substantially larger and more flexible guest molecules. (b) Optimal binding-conformation predictions for guests encapsulated within the $Pd_6L_4$ cages. Experimental structures are shown in yellow and *DeepHostGuest* predictions in magenta. Across hydrophobic, amphiphilic, and water-soluble guests, the model accurately reproduces experimental binding modes in most cases, including both rigid and flexible molecules with multiple rotatable bonds. RMSD values are indicated for each system. Hosts in predicted structures and hydrogen atoms are omitted for clarity.

This out-of-distribution generalization demonstrates that *DeepHostGuest* has learned transferable physical principles of molecular recognition rather than memorizing training examples. The model correctly interprets electrostatic and hydrophobic landscapes defined by host ESP surfaces, positioning even unfamiliar, flexible molecules according to fundamental recognition principles. This capability dramatically expands the framework's practical utility beyond its training domain.

**Integration with Free Energy Calculations**

Having established that *DeepHostGuest* predicts binding conformations with high fidelity, we next assessed whether these predicted structures provide quantitative information about binding affinity—the goal for rational design. We compiled approximately 70 experimental binding free energies ($\Delta G$ values) from literature involving diverse host families (cucurbiturils, pillararenes, organic cages, metal–organic cages) with experimental $\Delta G$ ranging from −1.93 kcal·mol$^{-1}$ (weak binding) to −13.25 kcal·mol$^{-1}$ (strong binding), covering the comprehensive spectrum of supramolecular interactions (see Supporting Figure S12). Experimental $\Delta G$ values were determined via multiple complementary techniques such as isothermal titration calorimetry (ITC) for solution-phase titrations and nuclear magnetic resonance (NMR) for equilibrium analysis. This methodological diversity introduces measurement uncertainty but also provides realistic coverage of experimental conditions.

DFT-calculated binding free energies based on *DeepHostGuest*-predicted structures correlate well with experimental values ($R^2$=0.72; mean absolute error, MAE = 1.74 kcal·mol$^{-1}$; root mean square error, RMSE = 2.14 kcal·mol$^{-1}$; Figure 4). This correlation magnitude is notable: it demonstrates that geometric accuracy (RMSD values) translates into energetic validity. Notably, the model distinguishes strong from weak binding with high reliability. High-affinity interactions are predominantly concentrated in CB[7]-based complexes, consistent with experimental observations and literature understanding that CB[7]'s larger cavity and appropriate polarity favor strong binding. Structural analysis of representative CB[7] complexes shows that cationic guest moieties preferentially locate near carbonyl portal oxygens, while hydrophobic fragments bury within the cavity interior—binding motifs that closely match established crystallographic patterns and empirical models from the supramolecular literature. For systems with available experimental crystal structures (such as cofacial organic cages binding polycyclic aromatic hydrocarbons), *DeepHostGuest* reproduces experimental geometries with RMSD < 3 Å and correctly identifies dominant interaction types (π–π stacking between guest aromatics and cage frameworks; Supporting

Figure S13 in Section 5). These comparisons further validate that predicted conformations accurately reflect both geometric and energetic characteristics. The DFT calculation procedures are detailed in Supporting Information Section 5.

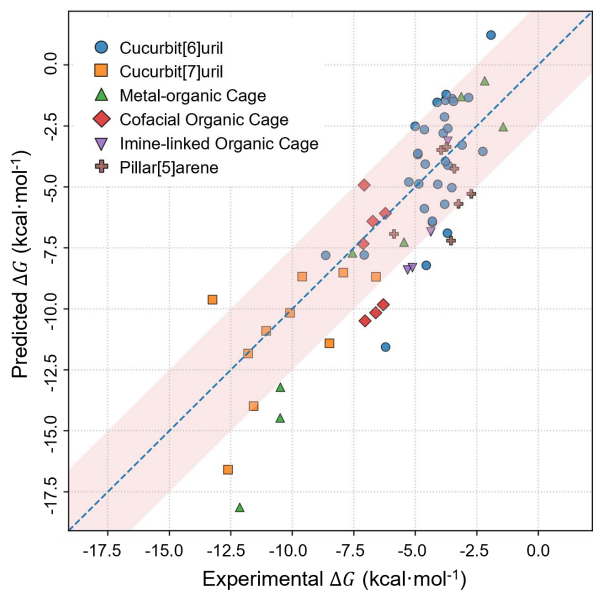

**Figure 4.** Correlation between experimental and calculated host–guest binding free energies. Comparison of experimentally measured binding free energies ($\Delta G$) and DFT-calculated $\Delta G$ values based on *DeepHostGuest*-predicted binding conformations for a diverse set of host–guest systems, including cucurbiturils (CB[6] and CB[7]), pillar[5]arenes, organic cages, and metal-organic cages. Each data point corresponds to one host–guest complex, colored and shaped according to host type. The dashed line denotes the linear regression between experimental and calculated values, and the shaded region depicts deviations within 2.5 kcal·mol$^{-1}$ from experiment. The correlation demonstrates that *DeepHostGuest*-predicted conformations provide a reliable structural basis for quantitative binding free-energy estimation across chemically diverse supramolecular systems.

The observed correlation scatter (±2.5 kcal·mol$^{-1}$) reflects realistic experimental limitations: different studies employ different solvents and measurement techniques, introducing systematic variations in reported $\Delta G$ values. The achieved correlation—0.72—is consistent with this inherent variability and suggests that *DeepHostGuest* predictions are robust to such variations.

These results demonstrate that *DeepHostGuest* provides structurally accurate and energetically meaningful conformations that enable reliable discrimination between strong and weak host–guest interactions. Building on this validation, we are expanding the experimental dataset to establish systematic structure-function relationship models, providing a scalable theoretical framework for the rational design and optimization of supramolecular host–guest systems.

**Implications for Supramolecular Design**

Building upon the small-scale dataset for DFT calculations, we expanded and constructed a larger host–guest database that integrates experimentally measured binding free energies, *DeepHostGuest*-predicted binding conformations, and multi-level physicochemical descriptors. This integrated dataset provides a quantitative foundation for systematically uncovering structure–activity relationships governing supramolecular recognition.

The database comprises 876 experimentally determined binding free energies spanning 34 distinct host architectures, including cucurbiturils, cyclodextrins, pillararenes, organic cages, and metal–organic cages, thereby covering a broader range of representative supramolecular systems (see Supporting Figure S14). For each host–guest pair, we first employed *DeepHostGuest* to generate the corresponding binding conformation, followed by the extraction of physicochemical descriptors at three distinct levels: host-level descriptors (polarity, surface properties, cavity geometry, MACCSKeys[36]), guest-level descriptors (hydrophobicity, size, surface properties, MACCSKeys), and complex-level descriptors (interaction fingerprints capturing various non-covalent interactions, and host–guest size complementarity), yielding a total of 19 feature types per complex (see Supporting Table S2 for the full list of extracted features).

Automated feature selection, model training, hyperparameter optimization, and ensemble learning were performed using *AutoGluon*. [37] *AutoGluon* evaluates tens of candidate models (gradient boosting, random forests, neural networks, and combinations thereof), performs cross-validation to estimate generalization performance, and selects the best-performing ensemble. The resulting predictive model exhibits strong cross-validated performance across a binding free energy range of −0.65 to −16.36 kcal·mol$^{-1}$, with a coefficient of determination $R^2$ = 0.80, MAE of 0.93 kcal·mol$^{-1}$, and RMSE of 1.18 kcal·mol$^{-1}$ (Figure 5a; Supporting Table S3). These results demonstrate that *DeepHostGuest*-derived conformations provide a robust structural basis for quantitative binding free energy prediction across chemically diverse host–guest systems. See Supporting Information Section 6 for dataset construction, feature extraction, and *AutoGluon*-based ML model training.

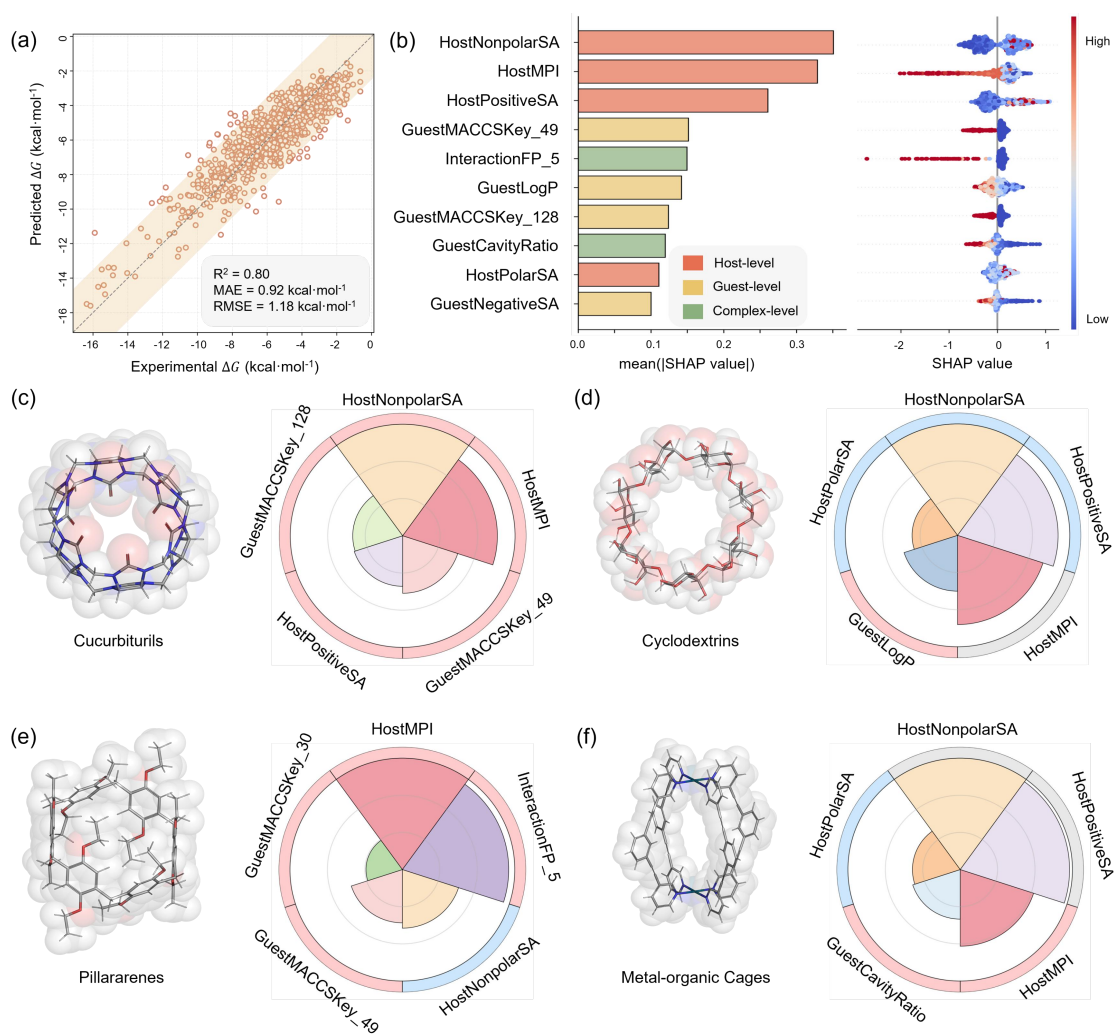

**Figure 5.** Structure–property relationship analysis of host–guest binding free energies enabled by *DeepHostGuest*. (a) Correlation between predicted and experimental binding free energies (*ΔG*) for 876 host–guest complexes spanning 34 host architectures; the shaded region indicates deviations within ± 2.5 kcal·mol$^{-1}$. (b) SHAP feature-importance ranking and beeswarm summary across the full dataset, showing that binding affinity is jointly governed by host polarity descriptors, guest chemical and structural features, and complex-level interaction and size-matching descriptors. Negative SHAP values correspond to features that enhance binding (more negative *ΔG*). (c–f) Top-ranked SHAP features for representative host families, revealing host-dependent structure–property relationships. Sector areas denote feature importance, and color encodes contribution trends (red: stronger binding with increasing feature value; blue: weaker binding; grey: non-monotonic). (c) Cucurbiturils benefit from increased inner-cavity nonpolar surface area. (d) Cyclodextrins show non-monotonic dependence on cavity size, with α- and β-cyclodextrins outperforming γ-cyclodextrin. (e) Pillararenes exhibit strong electrostatically driven binding for cationic guests. (f) Metal–organic cages combine host polarity, hydrophobicity, and size complementarity to stabilize encapsulated guests.

To elucidate the physicochemical origins of these predictions, we applied SHAP analysis[38] to quantify the contribution of individual features to binding free energy ($\Delta G$) (Figure 5b). SHAP analysis reveals that binding strength is governed by synergistic interplay between three feature categories: (1) host polarity descriptors (molecular polarity index, MPI, which measures charge distribution heterogeneity; nonpolar surface area; positive surface area), (2) guest chemical and structural features (octanol–water partition coefficient, logP, reflecting hydrophobicity; MACCSKeys capturing atom types or functional groups), and (3) complex-level interaction and size-matching descriptors (interaction fingerprints quantifying hydrogen bonds and cationic contacts; guest-volume-to-cavity-volume ratio). Mechanistically, higher host polarity (higher MPI) promotes electrostatic interactions and hydrogen bonding, enabling stronger binding. Guest hydrophobicity (higher logP) and appropriate size complementarity (GuestCavityRatio) stabilize complexes by promoting desolvation and maximizing van der Waals contact. Reduced positive surface area combined with charged guest features and cationic interaction fingerprints highlights the critical role of electrostatic complementarity in binding strength. In contrast, host nonpolar surface area exhibits non-monotonic contribution patterns, indicating its importance strongly depends on host class.

Host-specific SHAP analyses further reveal distinct binding regimes across supramolecular families (Figure 5c-5f; Supporting Figure S15 and Figure S16). In cucurbiturils systems, larger hydrophobic cavities enhance size complementarity and hydrophobic contact, enabling CB7 and CB8 to accommodate a broader range of high-affinity guests and achieve higher binding limits (Figure 5c).[39] Cyclodextrin binding is dominated by hydrophobic effects, where compact cavities with moderate external polarity outperform larger cavities; accordingly, α- and β-cyclodextrins exhibit stronger binding than γ-cyclodextrin (Figure 5d).[40] In pillararenes systems, synergistic effects between high host polarity and charged guests enable exceptionally strong binding exceeding −15 kcal·mol$^{-1}$, while guest heteroatoms further stabilize complexes via hydrogen bonding (Figure 5e).[41] Metal–organic cages display a distinct balance in which strong host polarization, appropriate cavity matching, and moderately hydrophobic environments jointly enhance binding (Figure 5f).[42]

These results establish *DeepHostGuest* as an integrated platform linking structural prediction with quantitative supramolecular design. By unifying accurate conformation prediction, binding free-energy estimation, and interpretable structure–activity analysis, the framework enables several key advances. First, it renders high-throughput virtual screening practical, allowing large host–guest libraries to be evaluated *in silico* prior to experiment. Second, supramolecular optimization becomes data-driven: SHAP-derived design rules provide actionable guidance for targeted modification of hosts and guests, replacing empirical trial-and-error. Third, *DeepHostGuest* expands access to challenging regions of chemical space, including metal–organic cages and unconventional supramolecular architectures, without additional parameterization. Finally, interpretable predictions deliver mechanistic insight, revealing why binding occurs rather than merely whether it occurs, thereby supporting the rational design of functional recognition systems. Looking ahead, expanding structural coverage, incorporating environmental effects, and integrating active-learning workflows will further enhance predictive accuracy and adaptability. Together, these developments position *DeepHostGuest* as a foundation for advancing supramolecular chemistry from empirical discovery towards a predictive, data-driven design paradigm.

**Conclusion**

This work establishes data-driven design as a practical framework for supramolecular host–guest recognition. *DeepHostGuest* demonstrates that GDL can extract transferable principles of molecular recognition directly from experimental structures, enabling accurate prediction of binding conformations across diverse supramolecular architectures without system-specific parameterization. The framework achieves high-fidelity structure prediction, outperforming classical docking and generalizing robustly to unfamiliar systems, including crystalline sponge platforms hosting large, flexible amphiphilic guests. This performance indicates that *DeepHostGuest* captures fundamental physicochemical features of recognition rather than memorizing specific chemistries. Importantly, *DeepHostGuest*-predicted structures provide a reliable foundation for quantitative binding free-energy estimation. DFT-calculated affinities correlate well with experiment, enabling systematic structure–activity analysis across 876 host–guest complexes spanning 33 host families. Interpretable analysis reveals cooperative design principles linking host electrostatics, guest hydrophobicity, and geometric complementarity, with distinct regimes across different supramolecular classes. SHAP-based attribution converts these correlations into actionable design rules for targeted host and guest optimization.

This framework fundamentally reframes supramolecular discovery from an empirical trial-and-error discipline towards a predictive, data-driven science. Where traditional approaches require extensive synthesis and screening to discover functional hosts and identify optimal guests, *DeepHostGuest* enables rational optimization through computational pre-screening, dramatically reducing the experimental burden. The ability to predict both binding geometry and affinity enables high-throughput virtual screening of unprecedented scale—evaluating thousands of candidate host–guest combinations *in silico* before experimental synthesis. This computational leverage is especially valuable for optimizing functional systems such as molecular containers for catalysis, separation, and drug delivery, where binding strength, selectivity, and cavity environment are precisely optimized.

Beyond supramolecular chemistry, *DeepHostGuest* exemplifies how GDL can solve inverse design problems in molecular science. The approach of learning scoring functions from experimental structures, validating predictions through independent physical calculations, and extracting interpretable design rules offers a reproducible methodology for fields ranging from crystal engineering to materials discovery. As experimental datasets and computational capabilities expand, this methodology will increasingly enable chemists to transition from optimizing systems at the margins of chemical space towards deliberately engineering molecular systems with unprecedented selectivity and function. In essence, *DeepHostGuest* achieves what synthetic supramolecular chemistry has long sought: a predictive framework that, like protein docking predictions, transforms molecular design from craft towards quantitative science. By establishing that non-covalent molecular recognition can be learned, predicted, and rationally optimized, this work provides both immediate tools for supramolecular researchers and a conceptual foundation for data-driven molecular design across chemistry.

## Supporting Information

The experimental section and further details, as well as tables and graphs, are presented in the supporting information.

## Acknowledgements

S.J. acknowledges financial support from the National Natural Science Foundation of China (22471169) and the Natural Science Foundation of Shanghai (24ZR1450900). S.J. also acknowledges the ShanghaiTech University Startup Fund to support this work. The AI-driven experiments, simulations, and model training were performed on the robotic AI-Scientist platform of the Chinese Academy of Sciences (CAS).

## Conflict of Interests

The authors declare no conflict of interest.

## Data Availability Statement

The curated host-guest structural dataset and experimental binding free energy dataset are available on https://zenodo.org/records/18222349. Code for *DeepHostGuest* model training, conformational prediction, binding free energy prediction, and SHAP analysis are available at https://github.com/Chemwzd/DeepHostGuest. Crystallographic data were obtained from the Cambridge Structural Database under appropriate licensing.

**Keywords:** Host-guest Chemistry • Molecular Recognition • Geometric Deep Learning • Structure–property Relationship Analysis